\documentclass[prl,amsmath,amssymb,twocolumn]{revtex4-1}
\usepackage[normalem]{ulem}
\usepackage{amsmath,amssymb}
\usepackage[bookmarks=true,colorlinks,linkcolor=blue,urlcolor=blue,citecolor=blue]{hyperref}
\usepackage{graphicx,amsmath,amsfonts}
\usepackage[usenames]{color}
\usepackage{epstopdf}
\usepackage{verbatim}
\usepackage{amsthm}
\usepackage{relsize}

\newcommand{\be}{\begin{equation}}
\newcommand{\beq}{\begin{eqnarray}}
\newcommand{\eeq}{\end{eqnarray}}

\def \tr{{\mbox{tr~}}}

\def \ua{{\uparrow}}
\def \da{{\downarrow}}
\def \be{\begin{equation}}
\def \ee{\end{equation}}
\def \ba{\begin{array}}
\def \ea{\end{array}}
\def \bea{\begin{eqnarray}}
\def \eea{\end{eqnarray}}
\def \nn{\nonumber}
\def \half{\frac{1}{2}}

\def \W{{\Omega}}

\def \a{{\alpha}}

\def \D{{\Delta}}
\def \d{{\delta}}

\def \G{{\Gamma}}

\def \av#1{{\langle#1\rangle}}
\def \ket#1{{\,|\,#1\,\rangle\,}}
\def \bra#1{{\,\langle\,#1\,|\,}}

\begin{document}

\title{Dynamical quantum phase transitions in random spin chains}
\author{Ronen Vosk$^1$ and Ehud Altman$^{1,2}$ \\{\small $^1$\em Department of Condensed Matter Physics, Weizmann Institute of Science, Rehovot 76100, Israel}\\ {\small $^2$\em Department of Physics, University of California, Berkeley, California 94720, USA}}
\begin{abstract}
Quantum systems can exhibit a great deal of universality at low temperature due to the structure of ground states and the critical points separating distinct states. On the other hand, quantum time evolution of the same systems involves all energies and it is therefore thought to be much harder, if at all possible, to have sharp transitions in the dynamics.
In this paper we show that phase transitions characterized by universal singularities do occur in the time evolution of random spin chains. The sharpness of the transitions and integrity of the phases owes to many-body localization, which prevents thermalization in these systems.
Using a renormalization group approach, we solve the time evolution of random Ising spin chains with generic interactions starting from initial states of arbitrary energy. As a function of the Hamiltonian parameters, the system is tuned through a dynamical transition, similar to the ground state critical point, at which the local spin correlations establish true long range temporal order.
In the state with dominant transverse field, a spin that starts in an up state loses its orientation with time, while in the "ordered" state it never does. As in ground state quantum phase transitions, the dynamical transition has unique signatures in the entanglement properties of the system.  When the system is initialized in a product state the entanglement entropy grows as $log(t)$ in the two "phases", while at the critical point it grows as $log^\a(t)$, with $\a$ a universal number. This universal entanglement growth requires generic ("integrability breaking") interactions to be added to the pure transverse field Ising model.
\end{abstract}
\maketitle

Closed systems evolving with Hamiltonian dynamics, are commonly thought to settle to a thermal equilibrium consistent with the energy density in the initial state. Any sharp transition associated with the long time behavior of observables must in this case correspond to classical thermal phase transitions in the established thermal ensemble. Accordingly, in one dimension, where thermal transitions do not occur, dynamical transitions are not expected either.

But systems with strong disorder may behave differently. Anderson conjectured already in his original paper on localization, that {\em closed} systems of interacting particles or spins with sufficiently strong disorder would fail to equilibrate\cite{Anderson1958}.
Recently, Basko et. al. \cite{Basko2006} gave new arguments to revive this idea of many-body localization, which has since received further support from theory and numerics\cite{Oganesyan2007,Pal2010,Znidaric2008,Bardarson2012,Vosk2013}. An important point for our discussion is that localized eigenstates, even at macroscopic energies, are akin to quantum ground states in their entanglement properties\cite{Huse2013,Serbyn2013a,Bauer2013}. In particular, it was pointed out in Ref. \onlinecite{Huse2013}, that localized eigenstates can sustain long range order and undergo phase transitions that would not occur in a finite temperature equilibrium ensemble. What are the observable universal signatures of such transitions?

In this paper we develop a theory of a critical point separating distinct dynamical states in the time evolution of random Ising spin chains with generic interactions
\be
H=\sum_i \left[ J^z_i  S^z_i S^z_{i+1}+h_iS^x_i + J^x_iS^x_iS^x_{i+1} +\ldots\right].
\label{eq:model}
\ee
Here $J^z_i$, $h_i$ and $J^x_i$ are uncorrelated random variables and $\dots$ represents other possible interaction terms that respect the $Z_2$ symmetry of the model.
For simplicity we take the distributions of coupling constants to be symmetric around zero.
Without the last term, $J^x_i$, the hamiltonian can be mapped to a system of non-interacting Fermions. We include weak "interactions" ($J^x_i\ll J^z_i,h_i$) to make the system generic.
\begin{figure}[t]
 \centerline{\resizebox{0.9\linewidth}{!}{\includegraphics{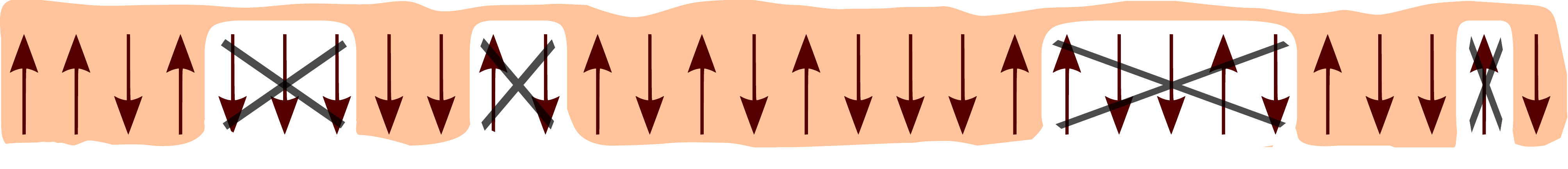}}}
 \caption{Schematic depiction of the spin glass phase. An infinite cluster is formed in the course of the RG, stoping the decay of spin correlations. }
 \label{fig:cluster}
\end{figure}

The transverse field Ising model (\ref{eq:model}) undergoes a ground state quantum phase transition controlled by an infinite randomness fixed point \cite{Fisher1994}. The transition separates between a quantum paramagnet obtained when the transverse field is the dominant coupling and a spin ordered state established when the Ising coupling $J^z$ is dominant. Recently, it was pointed out that this transition can also occur in eigenstates with arbitrarily high energy, provided that the system is in the many-body localized phase\cite{Huse2013}. Here we develop a theory of the non-equilibrium transition, focusing on the universal singular effects it has on the time evolution of the system in presence of generic interactions.

We shall describe the time evolution of the system starting from initial states of arbitrarily high energy. Specifically, we take random Ising configurations of the spins in the $S^z$ basis, such as $\ket{\psi_{in}}=\ket{\ua\ua\da\ua,\ldots\da\da\ua}$.  The theoretical analysis relies on the strong disorder real space RG approach (SDRG) \cite{Dasgupta1980,Fisher1992}, which we recently extended to address the quantum time evolution of random systems\cite{Vosk2013}. The properties of the transition are elucidated by tracking the time evolution of two quantities: spin correlations and entanglement entropy.

First, we show that the spin auto correlation function $C_z(t)=\bra{\psi_{in}}S^z_i(t)S^z_i(0)\ket{\psi_{in}}$
decays as a power-law in the paramagnetic phase, whereas it saturates to a positive constant in the spin (glass) ordered phase. Hence the operators $S^z_i$ have an overlap with integrals of motion in the ordered phase.
Second, the growth of the entanglement entropy between the two halves of the system depends crucially on whether the system is interacting or not. For generic interactions we show that the entropy grows as $\log(t)$ at long times in the two phases, whereas it grows as $[\log(t)]^\a$ at the critical point, with $\a$ a universal exponent greater than $1$. Enhanced entanglement at the critical point is a hallmark of ground state quantum criticality. Here the concept is extended to the entire spectrum and, correspondingly, to the dynamics at high energy density.

\noindent
{\bf Real space RG --} The approach we use to describe the time evolution of the random chain at long times
was presented in Ref. [\onlinecite{Vosk2013}]. The scheme is an extension of the SDRG method originally formulated to focus on long distance ground state correlations\cite{Dasgupta1980}, to the description of the dynamics on long time scales.

Let us review the basic idea of the scheme. The quantum evolution at the shortest times is governed by the largest couplings in the Hamiltonian, which we define as the cutoff scale $\Omega$. Spins, affected by much smaller couplings are essentially frozen on that timescale. The dynamics of these slow degrees of freedom at longer times can be described by an effective evolution operator that does not contain the high frequency scale $\Omega$. Technically we derive the effective evolution operator perturbatively in the coupling between the strongly coupled spins and their neighbors. The effective Hamiltonian has the same form as the original Hamiltonian, with renormalized couplings. Repeating the scheme leads to a flow of the distributions of the couplings which govern the evolution at increasingly long times $t=\hbar/\Omega$.

The two types of RG steps we need to consider are: (i) the case of a large coupling $J^z_i=\W$, and (ii) the case of a large transverse field $h_i=\W$. In case (i) the two spins coupled by the large Ising interaction can only flip collectively as a slow degree of freedom and we therefore join them to a new effective spin. The interactions acting on the new effective spin are as follows. The effective transverse field on it is $h_n= \eta h_1 h_{2}/\W+J^x_{12}$, where we denoted the two constituent spins by $1,2$ and $\eta=1\,(-1)$ if spins 1 and 2 are aligned (anti-aligned). The Ising interactions connecting the new spin to its nearest neighbors are unchanged up to a sign $J^z_{L}\to J^z_{L}$ and $J^z_{R}\to \eta J^z_{R}$ , while the $J_x$ interactions are reduced to $\tilde{J}^x_{L,R}=\eta J^x_{L,R} h_{2,1}/\Omega$. A smaller three spin interaction is also generated $\eta(J^x_L J^x_R/\W) S^x_L S^x_n S^x_R$ in a second order process. In addition, the transverse fields on the spins to the left and right of the new spin are slightly renormalized ${h}_{L,R}\to h_{L,R}+\eta J^x_{L,R} h_{1,2}/\W$. Note that because of the random sign of couplings at the outset, the signs of the generated interaction will be unimportant.

In case (ii) we have a large transverse field $h_n=\Omega$ on spin $n$. The effective Hamiltonian (in the interaction picture) for the slow evolution of the spins in that vicinity is
\be
H^{I}_{eff}={2J^z_L J^z_R\over \Omega} S^z_L S^x_n S^z_R+\sum_{\a=R,L}(J^x_\a S^x_n +h_\a)S^x_\a.
\ee
Since $S^x_n$ commutes with $H^I_\text{eff}$ we can take it as a number $\pm\half$, which depends on the spin projection along $\hat x$. The spin $n$ is eliminated at the expense of having a different effective hamiltonian $H^{\pm}_\text{eff}$ operating on initial states with the spin $n$ oriented along the positive or negative $\hat x$ axis. This hamiltonian includes an Ising interaction between the left and right neighbors of $n$: $\tilde{J}^z=\pm (J^z_LJ^z_R/\W) $ and a transverse field $\tilde{h}_{L,R}=h_{L,R}\pm\half J^x_{L,R}$.

Note that  in case (ii), at this level of approximation, no $J^x$ term is produced between the left and right neighbors. However, as we argue below, such interactions are produced in the more generic Hamiltonian that emerges at intermediate scales after the RG has progressed for some time. Recall that  type (i) decimations produce three spin interactions terms of the form $S^x_{i-1}S^x_i S^x_{i+1}$. This additional term, in turn, may give rise to a four spin interaction term. In fact, as we show in the SI, all strings of $n$ spin interactions of the form $S^x_1 S^x_{2}\ldots S^x_{n}$ are produced at some stage of the RG, with coefficients that decay exponentially with the length of the string. The most important contribution of the strings is that if a spin is decimated at the center of a three spin interaction, an effective $J^x$ interaction is produced $(J^x_L J^x_R/2\W)S^x_L S^x_R$.

At this point we can discuss the nature of the RG flow, from which we later derive the main results.
Broadly the flow implies two phases:  one dominated by the transverse field, and another, dominated by the Ising interaction, which we will call "glass", separated by an infinite randomness critical point.

\begin{figure}[t]
 \centerline{\resizebox{0.9\linewidth}{!}{\includegraphics{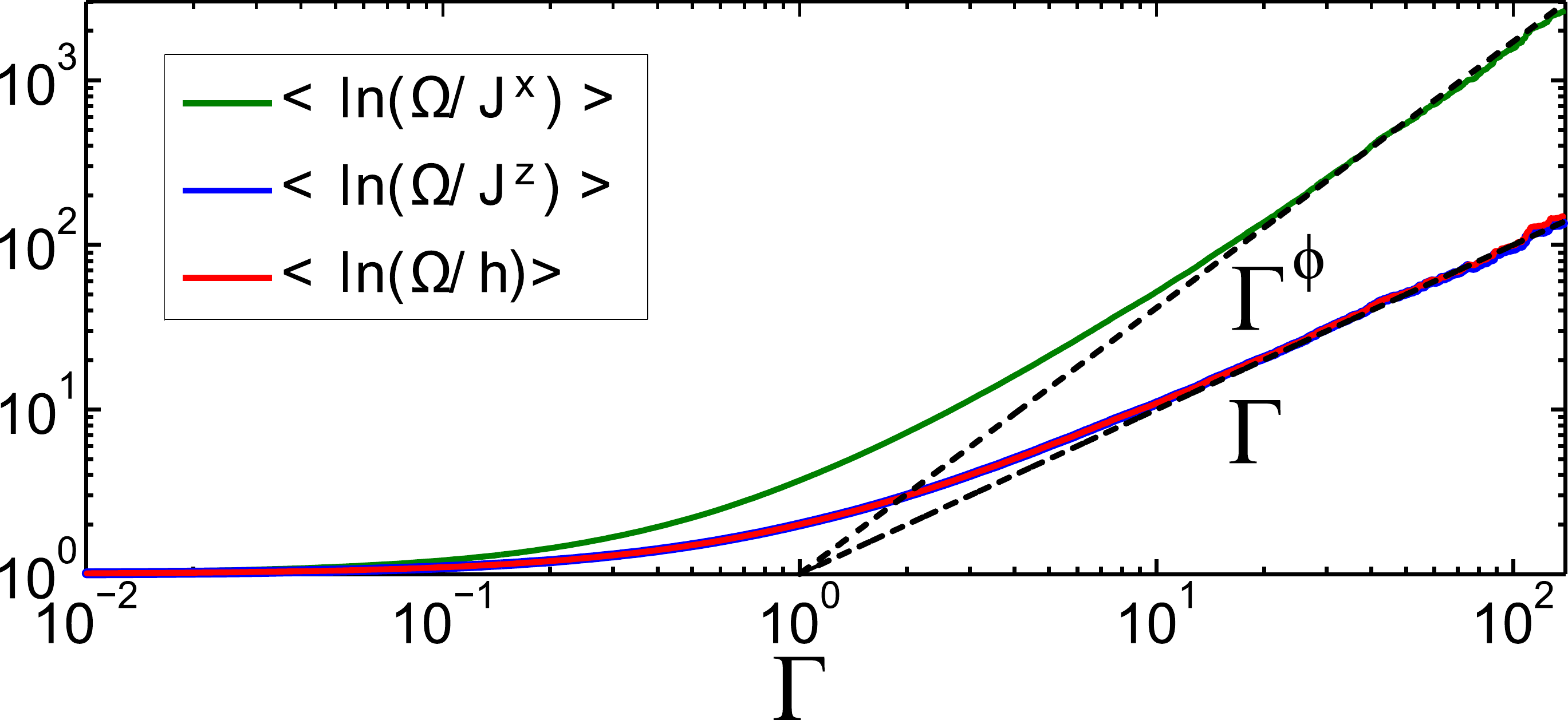}}}
 \caption{The flow of the averaged scaling variables with $\G$ at criticality. The interaction term $\av{\ln(\W/J^x)}$ scales asymptotically as $\G^\phi$ while the other variables $\av{\ln(\W/h)}$ and $\av{\ln(\W_0/J^z)}$ scale as $\G$.}
 \label{fig:critical_flow}
\end{figure}

Fig.  \ref{fig:critical_flow} shows the flow of the scaling variables associated with the different couplings  as a function of the RG scale $\G=\ln(\W_0/\W)$ at criticality. The numerical results were obtained by operating the RG rules on a long chain of $10^6$ sites.
We see that $\av{\ln(\W_0/J^z)}, \av{\ln(\W/h)}$ scale linearly with $\G$, just as expected from the TFIM without the interaction\cite{Fisher1992}, while  $\av{\ln(\W/J^x)}\sim \G^\phi$, with $\phi= (1+\sqrt{5})/2$ the golden ratio.
As a result, the typical value of the interaction decays as $J^x_{typ}\sim \exp(-\G^\phi)$, much faster than the other couplings and does not feed-back onto their flow. This ensures the stability of the infinite randomness dynamical fixed point to interactions introduced within the SDRG scheme.

The interactions have another effect, on the dynamics only, that is not captured by the SDRG scheme. They can destabilize the localized phase by mediating resonances between modes that oscillate with similar frequencies $\W\pm\d\W$ at remote sites on the chain.
In Ref. \cite{Vosk2013} we argued that such resonances do not proliferate and are therefore irrelevant for sufficiently strong disorder. This is due to the the $1/L$ dependence of the energy mismatch of potential resonant sites found within a range $L$ compared to the exponential decay with $L$ of the effective interaction between such sites.

The fact that the interactions, embodied by $J^x$, are irrelevant does not mean they can be completely neglected. We will show that in the course of their flow to zero the interactions have a dramatic effect on the growth of the entanglement entropy in the system, giving rise to the universal $\log(t)$ increase in the two phases and $[\log(t)]^\a$ at the critical point.

\noindent
{\bf Spin decay --}
The initial state of the time evolution is a product state of spins with well defined $S^z_i$ on every site. A simple  question we can ask is how the local spin orientation is disordered at long times. This is quantified by the spin auto-correlation time $C_z(t)=\bra{\psi_{in}}S^z_i(t)S^z_i(0)\ket{\psi_{in}}=\av{S^z_i(t)_i}\av{S^z_i(0)}$, or the disorder average of this correlation function $\overline{C_z(t)}$.

Within the SDRG scheme described above a spin maintains its orientation when it is being joined into larger clusters. Only when the cluster it belongs to is decimated due to a large transverse field does the test spin lose its orientation. Therefore the average moment of a spin is directly related to the probability it had not been decimated by the time of measurement. This is given by $\half (N_r/N_0)$, where $N_r$ is the number of original spins, which belong to undecimated clusters and $N_0$ is the total number of spins.

Exactly the same ratio $m\equiv N_r/N_0$ enters the calculation of the ground state magnetization density in the standard SDRG scheme\cite{Fisher1992,Fisher1995}. Hence, we can read off the results and translate them to the time evolution. In the paramagnetic phase we have $m\sim \G\exp(-2\D\G)$, where $\D$ is the detuning from criticality (See also Ref. \cite{Igloi2005}). In our case $\G=\ln(\W_0 t)$ and we see that the spin autocorrelation decays to zero as $\overline{C_z(t)}\sim \ln(\W_0 t)/(\W_0 t)^{2\D}$. 
At criticality ($\D=0$), the power law reverts to logarithmic decay $\overline{C_z(t)}\sim  [1/\ln(\W_0t)]^{2-\phi}$. Finally, in the "glass" phase, dominated by $J^z$, one cluster grows to include a finite fraction of the original spins and is never decimated. Hence in an infinite system the autocorrelation function saturates to a positive constant at long times. The saturation value of the autocorrelation function is the order parameter of the "glass" phase.
Again using the correspondence with the ground state magnetization density we can determine
the onset of the order parameter at the critical point as $\overline{C_z(\infty)}\sim |\D|^{2-\phi}$.

\noindent
{\bf Growth of entanglement entropy --} We now turn to discuss the entanglement entropy $S = -\tr\rho_A \log_2\rho_A$ between the two halves of the chain $A$ and $B$. While the initial state is a non-entangled product state, correlations between the two halves are gradually generated in the course of time evolution. We can gain more information on the nature of the dynamical critical point and on the two phases from the rate at which entanglement is produced. In particular, we will see that the integrability breaking interactions have a dramatic effect on the growth of entanglement entropy, whereas they do not affect the local spin correlations discussed above.

Start with the non-interacting system, $J^x_i=0$. The process that generates correlations between the two sides in this case is the collective oscillation of clusters of spins that cross the interface. In the SDRG scheme this corresponds to a decimation of a spin by a large transverse field. When the decimated spin represents a cluster of original spins that cross the $AB$ interface, an order unity of entanglement entropy is added to the total count.  This is the same as the counting for the ground state entanglement\cite{Refael2004}. It leads to growth of entanglement as  $S\sim \ln\ln t$, as observed numerically \cite{Igloi2012}.

In a finite chain of length $L$ at criticality the entanglement growth is cutoff when all spins had been decimated. Infinite randomness scaling between length and time implies the saturation time $t_L\sim \W_0^{-1} e^{\sqrt L}$, leading to the entropy $S(L)\sim \ln L$. On the other hand, if the system is detuned by a small amount $\D$ from criticality it is dominated by the critical flow up to the emergent localization length $\xi\sim \D^{-\phi}$.  The entanglement entropy therefore grows initially as $\ln\ln t$, but eventually saturates to a finite value $S_\infty\sim \ln\xi$ even in an infinite system (see the SI for an explicit calculation).

In presence of interactions ($J^x_i\ne 0$) there is
a new process which contributes to the entanglement growth.
When a spin is decimated due to a large transverse field, the effective Hamiltonian acting on the nearby spins, $H_\text{eff}^\pm$ depends on the orientation of the decimated spin along the $x$-axis. Since the spin initially points along $z$, and therefore a superposition of projections on $x$, the two  distinct evolutions occur in parallel, thus producing entanglement after a delay time $t_{ent}\sim 1/J^x_\text{typ}$, set by $H^+_\text{eff}-H^-_\text{eff}$.

Let us compute the contribution of this process to the growth of entanglement entropy. A spin decimated at time $t_1$ near the $AB$ interface entangles with its neighbors by the time $t=t_1+t_{ent}$. The space between the decimated spin and its neighbors at $t_1$ contains many spins, that had already been decimated. These spins are associated with a smaller delay time, and by the observation time $t$ must also be fully entangled with each other. Hence, the entanglement entropy at time $t$ is the number of such spins $S(t)\sim \approx l_{\G_1} p_h(\G_1)$. Here $l_{\G_1}$ is the separation between the surviving spins at that stage of the RG in units of the original spins and $p_h(\G_1)$ is the fraction of  those spins that had been decimated by a large $h$. At long times $p_h$ is simply a constant equal to $1/2$ at criticality (see SI for details). The time-length scaling $l(\G)$ is well-known from ground state results\cite{Fisher1992}. But we must find the relation between $\G_1$ and $\G$.

The fact that $J^x$ decays rapidly in time ensures that $t_{ent}(t_1)=1/J^x_{typ}(t_1)\gg t_1$. Therefore $t\approx t_{ent}(t_1)$, or equivalently $\G\approx \G_1+ \av{\ln(\W_1/J_x)}\approx \av{\ln(\W_1/J_x)}$. The precise relation between $\G_1$ and $\G$ now depends on whether or not we are at the critical point.

At criticality Fig. \ref{fig:critical_flow} shows that $\av{\ln(\W_1/J_x)}\sim \G^\phi$ and therefore $\G_1\sim \G^{1/\phi}$. Plugging this into the expression for the entropy we have:
\be
S=\half l_{\G_1}\sim \half \G_1^2\sim \half \G^{2/\phi}\sim \half(\ln t)^{2/\phi}
\ee
Interestingly, the asymptotic time dependence  is the same as was obtained in the critical XXZ chain\cite{Vosk2013}.

Off-criticality $\av{\ln J^x/\W_1} \sim e^{2|\D| \G_1}$ in both phases (see SI) and therefore $2|\D|\G_1=\ln\G$. The length scale grows with the same exponential rate\cite{Fisher1995} $l_{\G_1} \sim e^{2|\D| \G_1}=\G$  Substituting into the expression for the entanglement entropy we have
\be
S^{\D\neq0}_{J^x\neq0} = p_h l_{\G_1} \sim p_h\G\sim p_h \ln t.
\ee
In both cases the interaction induced growth of the entanglement entropy begins after a delay $t_{d}\approx 1/J^x_0$, where $J^x_0$ is the typical value of the interaction at the outset.

The logarithmic growth of entanglement in generic localized phases has been observed in numerical simulations\cite{Chiara2006,Znidaric2008, Bardarson2012}. A heuristic argument for this behavior was recently given in Refs. \cite{Serbyn2013,Huse2013a}.

In a system of length $L$ the entanglement entropy, both on and off criticality,
 saturates to a value linear in $L$: $S(t)= p_h l_{\G_1}\to p_h \,L$. However this extensive entropy  does not imply thermalization of the system.
 since $p_h$, the fraction of large field decimations, is less than $1$, the saturation value of the entropy is smaller than the expected thermal entropy of 1 unit ($\log_2 2$) per spin (for generic initial states with energy in the middle of the many-body spectrum). In fact, $p_h$ increases monotonically as the system is tuned from the glass to the "paramagnetic" phase.

\noindent
{\bf Discussion --}  The absence of thermalization can be associated with the emergence of local conserved quantities, whose value is constrained by the initial state. In our case, these would be operators that involve $S^z_i $ since the spins have a well defined $z$ projection at the outset.

The fact that $\av{S^z_i}$ does not decay to zero in the "glass" phase implies that this operator is closely related to a true conserved quantity. Specifically, on sites $i$ that will eventually join the infinite cluster, $S^z_i$ has overlap of order $1$ with a conserved quantity $\tilde{S}^z_i$. As discussed above in relation to the decay of local spin correlations, these sites have a low density on the chain proportional to $\D^{2-\phi}$ near the critical point.
We can  then write the general form of these integrals of motion consistent with scaling  near the critical point as
\be
\tilde{S}^z_i= A S^z_i + B \exp\left(-C \D^\phi n\right) \hat{O}_{i,n},
\ee
where $O_{i,n}$ are strings of $n$ spin-$1/2$ operators and $A, B, C$ are constants of order $1$.
The decay of non local terms stems from the finite correlation length $\xi\sim 1/\D^\phi$ in the glass phase. In contrast to the spins on the infinite cluster, other spins have a finite, but exponentially suppressed overlap with true conserved quantities.

At the critical point and in the paramagnetic state $S^z_i$ is no longer a quasi-conserved operator.
However, there are other local conserved quantities. For example, consider a pair of spins belonging to the same cluster that is decimated due to a large transverse field. While each one of the spins is not conserved, their product $S^z_1 S^z_2$ is quasi-conserved in the same sense as above. Therefore the entropy does not reach the maximal thermal value in either phase. Note however, that deep in the paramagnetic phase almost all decimations are of a single spin and there are essentially no conserved quantities that contain $S^z_i$. In this limit the increase of the entropy is not constrained. This is consistent with the expression derived above $S_\infty =p_h L$, where deep in the paramagnetic phase $p_h\to 1$. Of course this does not mean that in this regime the system thermalizes for any generic initial conditions. Indeed all the $S^x_i$ are quasi-conserved quantities deep in the paramagnet.  If we took initial states in which the $S^x_i$ had definite values, the system would not thermalize.

\noindent
{\bf Conclusions --}
Using a real space RG method formulated in real time, we developed a theory of a dynamical quantum phase transition between distinct many-body localized phases of a quantum spin chain. The two phases are a simple paramagnet and a spin-glass are separated by an infinite randomness fixed point. The spin glass is characterized by long range temporal order in the spin, that is $\av{S^z_i(t)S^z_i(0)}$ saturates to a finite value. The saturation value onsets as $\D^{2-\phi}$ and serves as an order parameter of the dynamical phase. We note that a paper posted in parallel to this one explores a similar dynamical transition from the complementary perspective of dynamical response in a system prepared at equilibrium.\cite{Pekker2013}

Unlike the spin correlations, the growth of entanglement entropy following the quench is dramatically affected by interactions. While without interactions the entropy saturates in the two phases, the interactions lead to (delayed) logarithmic growth of the entanglement entropy. At criticality the $\ln t$ growth is enhanced to $\ln^{2/\phi} t$. It is important to note that the interaction induced entanglement does not involve energy transfer, but rather dephasing between localized modes centered at remote sites. However this process can have measurable consequences. For example, it will set the limit on the ability to protect quantum information for long times through localization.

In a finite system the entropy grows to an extensive value, but with smaller entropy density than it would reach in thermal equilibrium. An infinite set of emergent conserved quantities whose value is constrained by the initial state prevents the system from thermalizing. In the glass phase the local spin operators $S^z_i$ are quasi-integrals of motion. The scaling properties of the critical point can be used to write the general form of the true integrals of motion in its vicinity.

\noindent
{\bf Acknowledgements --}
We thank J. E. Moore, D. Huse, G. Refael, and D. Pekker for useful
discussions. This work was supported by the ISF, the
Minerva foundation, and the NSF Grant No. PHY11-
25915 during a visit of E. A. to the KITP-UCSB. E. A.
 acknowledges the hospitality of the Miller
institute of Basic research in Science.
\bibliographystyle{phd-url-notitle}
\bibliography{mbl}

\newpage

\appendix
\section{Supplementary Material}
\subsection{String interactions in the SDRG}
In this section we describe the generic interaction terms that we include in the Hamiltonian, and explain how they are naturally generated in the course of renormalization.

Suppose we have only nearest neighbor interactions at the outset of the flow.
As explained in the main text, an RG decimation due to large $J^z$ coupling the spins at sites $1,2$ gives rise to an  effective 3 spin interaction term
\be
{J^x_R J^x_L\over \Omega}S^x_{L} S^x_n S^x_{R}
\label{3spin}
\ee
where $S^\a_n$ is the new effective spin formed by joining  spins $1$ and $2$.
If later on a bond adjacent to those three spins is decimated due to a large $J^z$, a four spin interaction is generated. The same process continues to produce string interactions of all lengths.

Since string interactions are anyway generated we might as well include them from the beginning. However, we should do this in a way that is consistent with how the interaction terms are actually produced in the RG so that their form remains self similar in the course of renormalization. To do this we take our hint from the above three spin term. However, the way the coefficient is presented in (\ref{3spin}) is not very useful, because it is written in terms of the interactions acting in the original chain. Let us rewrite it in terms of the renormalized interactions that operate in the chain after the $J^z$ decimation had taken place. In the main text we found the  effective interaction terms coupling the new spin $n$ to the left and right $\tilde{J}^x_{L,R}= J^x_{L,R} h_{1,2}/ \Omega$, we also have the transverse field on the new spin $\tilde{h}_n\approx h_1 h_2/\Omega$ (to leading order in $J^x$). Plugging this into (\ref{3spin}) we obtain the coefficient of the three spin string  expressed in terms of the interactions on the renormalized chain
\be
J^x_{(3)}={J^x_R J^x_L\over \Omega}={\tilde{J}^x_L\tilde{J}^x_R \Omega \over h_1 h_2}\approx {\tilde{J}^x_L\tilde{J}^x_R\over \tilde{h}_n}
 \ee

Here and in what follows we use the bracketed subscripts, as in $J^x_{(m)}$ to denote the length of the string interaction, whereas unbracketed subscripts as in $J^x_i$ represent a site index.

Similarly, the longer interaction strings produced in the flow have the form
\be \label{eq:string}
H^i_{(m)} = \left({\prod_{l=i}^{i+m-2} J^x_{l}\over \prod_{l=i+1}^{i+m-2} h_l} \right)\left( \prod_{l=i}^{i+m-1}S^x_l\right)  ,
\ee
Note that the product of $J^x_l$ in the numerator is over all the interactions $J^x$ acting along the string whereas the product in the denominator is over the transverse fields acting on the spins in the string not including the first and last spins. Thus the coefficient of a string of length $m$ is of order $J^x(J_x/h)^{m-2}$. Since we assumed that initially $J^x\ll h$, the coefficients decay exponentially with the length of the string.
We include all strings of all lengths $m\ge3$

\begin{figure}
 \centerline{\resizebox{0.9\linewidth}{!}{\includegraphics{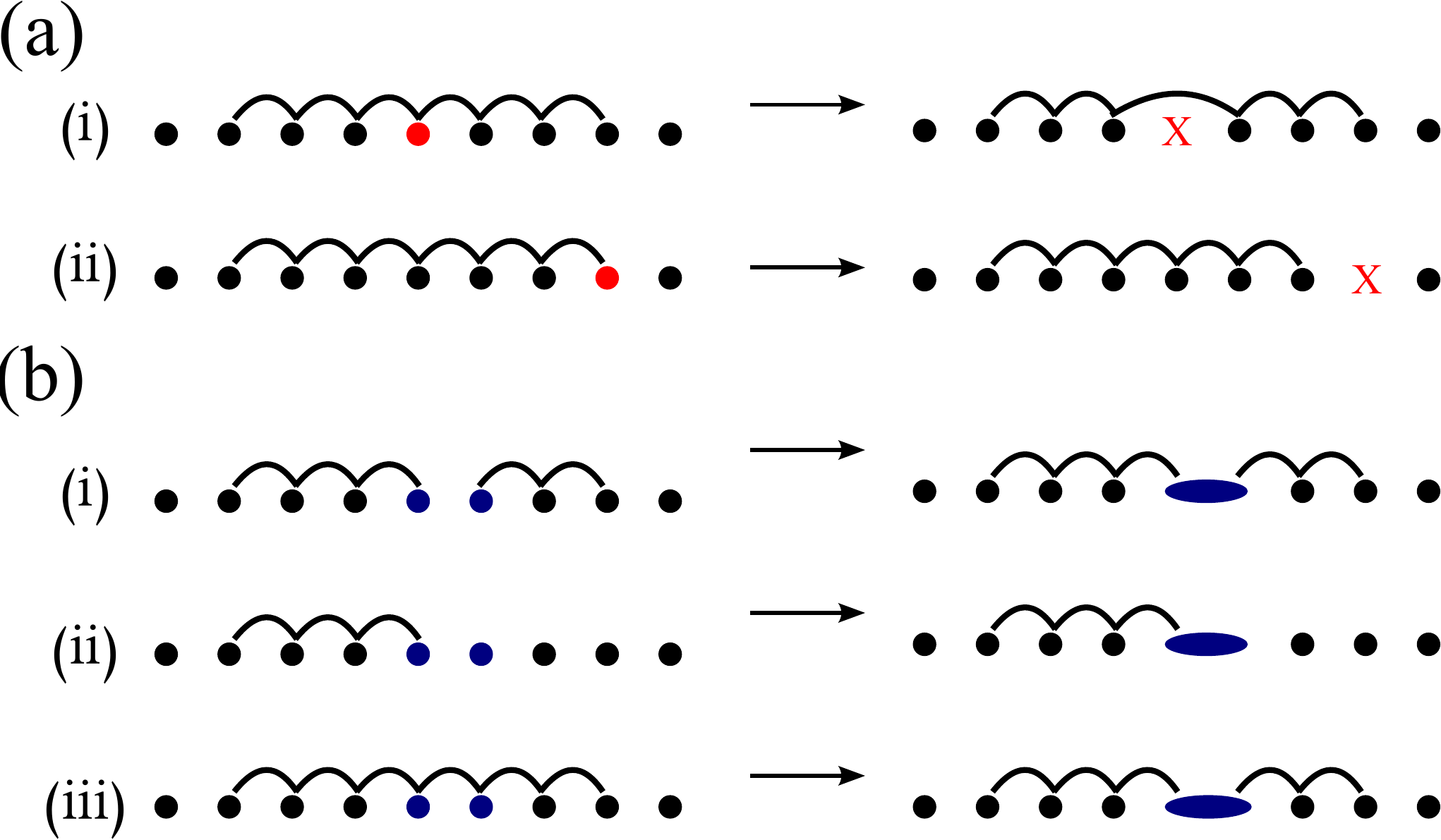}}}
 \caption{Decimation of strings. (a) Strong field decimation: (i) the decimated spin is in the bulk of the string or (ii) it is at the edge. (b) Strong bond decimation (i) joining of two strings that ended on the two sites. (ii) shortening of a string that includes only one of the two sites. (iii) shortening of a string that includes both sites.}
 \label{fig:strings}
\end{figure}

Let us verify that when such strings are included, the interactions produced in the RG decimations remain self similar with coefficients as proposed in Eq. (\ref{eq:string}) to leading order in $J^x$. Consider a decimation of a site $i$  due to a large transverse field $h_i=\W$.
All strings which contain this site will be shortened by $1$. Moreover the transverse field term and the string interaction commute. Therefore we may simply replace the decimated spin by a number $\pm\half$. Therefore the shortened string has exactly the same coefficient as the original string times a factor of $\pm\half$. But we should express this coefficient in terms of the renormalized interactions on the new string in order to verify that it matches the form (\ref{eq:string}). Here we should separate the discussion to two cases shown in Fig. \ref{fig:strings}(a): (i) the decimated spin is in the bulk of the string or (ii) it is at the edge.

The prototype of case (i) is decimation of the central spin $i$ in a three spin string. This leads to a {\em two spin} interaction $\tilde{J}^x_{i-1}S^x_{i-1}S^x_{i+1}$ with $\tilde{J}^x_{i-1}=J^x_{i-1} J^x_{i}/(2\Omega)$.   This interaction is a product of, but not contained among the string interactions defined in Eq. (\ref{eq:string}). Instead this process defines how an effective interaction is produced in a site decimation step.

Now consider a longer string with the decimated site $i$ ($h_i=\W$) somewhere in its bulk. We can re-express the coefficient of the shortened string in terms of the renormalized interactions to obtain exactly the form of (\ref{eq:string}):
\bea
\tilde{J}^{x}_{(m-1)}&=&{J^x_1 J^x_2\ldots J^x_{i-1} J^x_i\ldots J^x_{m-1}\over 2\,h_2\ldots \W\ldots h_{m-1}}\nn\\
&=&{J^x_1 J^x_2\ldots \tilde{J}^x_{i-1}  J^x_{i+1}\ldots J^x_{m-1}\over \,h_2\ldots h_{i-1} h_{i+1}\ldots  h_{m-1}}.
\eea

Now to case (ii), the decimated spin is the first (or last) spin in the chain.
The new shorter string which is formed is already contained in the Hamiltonian. Since the correction to the coefficient comes from a string of longer length, it is of higher order in $J^x$ than strings of the same length already present. We shall omit its contribution.

We turn to discuss formation of strings as part of the bond decimation step, i.e. when  $J_i^z = \W$ on a certain bond. We label the spins coupled by the the strong bond by $i$ and $i+1$. The processes to consider are shown in Fig. \ref{fig:strings}(b). They include (i) joining of two strings that ended on the two sites. (ii) decimation of a string that includes only one of the two sites. (iii) shortening of a string that includes both sites.

In process (i) the end spins of two strings are joined to a single new spin thus fusing the two strings. If the two original strings are of length $m$ and $n$, the newly formed string is of length $m+n-1$. The new string coupling is given by:
\bea
J^x_{(m+n-1)}&=&{J^x_{(m)} J^x_{(n)}\over \Omega}={J^x_1 J^x_2\ldots J^x_{m-1} J^x_{m+1}\ldots J^x_{m+n-1}\over \,h_2\ldots h_{m-1}\W \,h_{m+2} \ldots h_{m+n-1}}\nn\\
&=&{J^x_1 J^x_2\ldots \tilde{J}^x_{m-1} \tilde{J}^x_{m+1}\ldots J^x_{m+n-1}\over \,h_2\ldots h_{m-1}\tilde{h}_m \,h_{m+2} \ldots h_{m+n-1}}
\eea
which is exactly the right coefficient. In the last line we converted from the original couplings to the effective $\tilde{J}^x$ coupling of the new spin to its neighbors and the effective transverse field on the new spin $\tilde{h}_m$ using:
$\tilde{J}^x_{m- 1}=J^x_{m-1} h_m/\Omega$, $\tilde{J}^x_{m}=J^x_{m+1} h_m/\Omega$ and
$\tilde{h}_m= h_m h_{m+1}/\Omega$. Note that because two spins were joined there are no couplings $h_{m+1}$ and $J^x_m$.

In process (ii) the spin on site $m$, the right edge of a string of length $m$, is joined with the next spin $m+1$. The new string coupling is:
\be
\tilde{J}^x_{(m)}={J^x_{(m)} h_{m+1}\over \W}={J^x_1\ldots \tilde{J}^x_{m-1}\over h_2\ldots h_{m-1}}
\ee
exactly the correct form in terms of the renormalized local coupling constants following this RG step.

Finally process (iii) describes the shortening of a string of length $m$ due to merging two spins that belong to it to a single spin. This produces a string interaction with $m-1$ spins but with coefficient of order $\tilde{J}^x(\tilde{J}^x/\tilde{h})^{m-2}$, one order higher than strings of the same length produced by the other processes. We can therefore neglect the contribution of this process to the production of new strings.

We have thus shown by inspection that all the leading order string interactions generated in the course of renormalization conform exactly to the form (\ref{eq:string}) which is maintained throughout the flow.

\subsection{Entanglement Entropy evolution off-criticality in the non-interacting case ( $J^x=0$)}
As explained in the main text, to compute the evolution of the entanglement entropy in the absence of interactions we need to count the number of large field decimations that occur over the interface between the two halves of the system. Each decimation contributes a unit of entanglement entropy. This counting requires the solution of the flow equations, which we now briefly review.

Below we briefly review the elements of the solution to the RG flow, needed for the counting. The RG steps generate a flow  of the distributions of the couplings $R(\zeta;\G)$ and $P(\beta;\G)$, where $\zeta=\ln(\W_0/J^z)$ and $\beta = \ln(\W_0/h)$. The flow equations are solved by exponential distributions, $R(\zeta;\G) = R_0(\G) e^{-R_0(\G) \zeta}$ and $P(\beta;\G) = P_0(\G) e^{-P_0(\G) \beta}$.

The solution is written compactly in terms of the new variables $\D = \half(P_0-R_0)$ and $Y(\G) = \half(R_0+P_0)$. $\D$ is the detuning parameter from criticality, and is an invariant of the RG flow. The flow equation for $Y$ is solved by
\be \label{eq:flow_solution}
Y(\G) = \frac{\D {Y}_0+\D^2 \tanh \left(\D \G\right)}{\D +{Y}_0 \tanh \left(\D \G\right)},
\ee
where ${Y}_0 = Y(\G=0)$.

We can now begin to count the number of large transverse field decimations of clusters that contain the interface. At a given stage of the RG the interface may either separate two clusters or be contained within a cluster. If we start from the first situation we first need a large $J$ decimation, i.e. cluster formation over the interface. Entanglement will be produced when that cluster is decimated in a later stage. On the other hand after decimating a cluster that contains the interface we will be back to the original state with the interface between two clusters. Therefore entanglement is produced in two necessarily consecutive processes of cluster formation (large $J^z$) and then cluster decimation (large $h$).

When $\G$ is increased by $d\G$ we may write the contribution to the counting as the product of the total number of field decimations at the interface $P_0 d\G$ times the probability that the previous decimation was a bond decimation. The latter is the ratio of bond decimations to the total number of decimations: $R_0/(R_0+P_0)$, which should be evaluated at $\G-d\G$. However, since these functions are continuous we can simply evaluate them at $\G$.

Using the arguments above, the entanglement entropy is given by the integral
\be
S^{\D\neq0}_{J^x=0} \sim \int_0^\G \frac{R_0 P_0}{P_0+R_0} d\G.
\ee
Using \eqref{eq:flow_solution}, this integral can be calculated exactly to give
\be
S^{\D\neq0}_{J^x=0} (\G \rightarrow \infty) \sim \half \ln \left({Y}_0 / |\D|\right).
\ee
The saturation value of the entanglement entropy is independent of the length of the system $L$.

\subsection{Fraction of spins decimated by large transverse field}
In the main text we defined $p_h (\G)$, the ratio of the number of sites decimated by large field decimations up to time $\G$, to the total number of decimated sites. This fraction is used in the calculation of the entanglement entropy for non-zero interactions.  Here, we calculate it explicitly and show that it goes to a constant as $\G\rightarrow\infty$.

The first step is to calculate the numerator of $p_h (\G)$. The number of spins decimated by the $h$ rule when $\G$ is increased by $d\G$ is given by $N_0 P_0 n_\G$, where $n_\G$ is the fraction of remaining spins in the chain. Therefore, the total number of spins decimated by the $h$ rule up to scale $\G$ is given by
\be
N_0\int_0^\G d\G^\prime P_0(\G^\prime) n_{\G^\prime} =  \frac{N_0}{2} (1-n_\G) + \frac{N_0\D}{{Y}_0 + \D \coth (\D \G)}.
\ee
In the evaluation of the integral we used the solution for $n_\G$~\cite{Igloi2005}
\be
n_\G = \left[ \cosh \left(\D \G\right)+\frac{{Y}_0}{\D}\sinh \left(\D \G\right)\right]^{-2}.
\ee

To find $p_h (\G)$, we simply have to divide the above result by the total number of decimated spins, $N_0(1-n_\G)$. Since we are interested in asymptotically long times, we take the limit $\G\rightarrow \infty$, and use $n_{\G\rightarrow\infty} = 0$. We find the final result
\be
p_h (\G\rightarrow \infty)=\frac{1}{2}  + \frac{\D}{{Y}_0 + |\D|}
\ee
Note that $p_h (\G)$ goes to a constant at long times. Therefore, it only contributes an overall pre-factor in the entanglement entropy, and does not change the asymptotic time evolution.

\section{Scaling of the interactions off criticality}
\begin{figure}
 \centerline{\resizebox{0.9\linewidth}{!}{\includegraphics{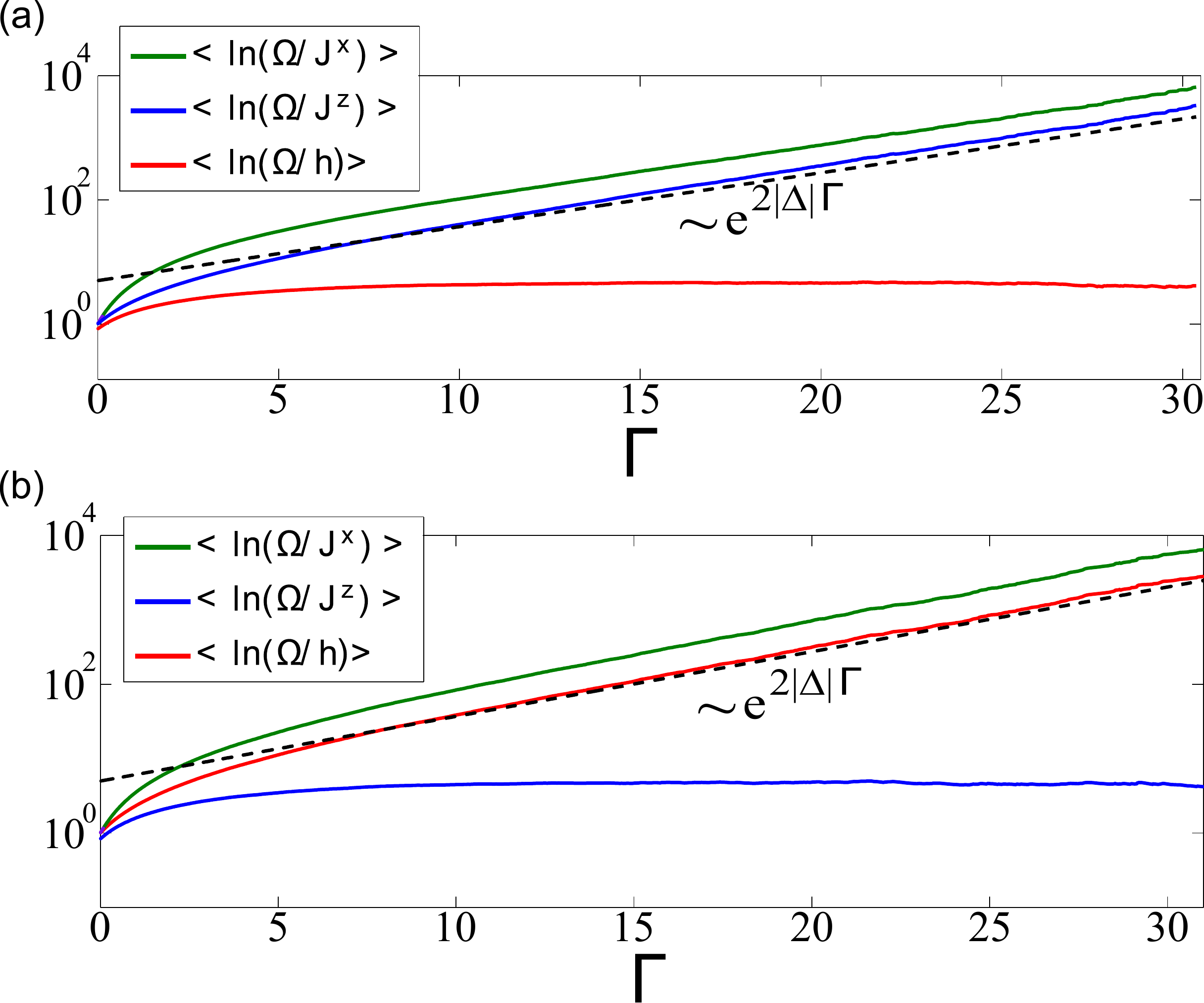}}}
 \caption{The flow of the averaged scaling variables with $\G$ off criticality. In (a) the system is in the paramagnetic phase with $\D=0.1$, and in (b) the system is in the spin glass phase with $\D=-0.1$. The interaction term $\av{\ln(\W/J^x)}$ scales asymptotically as $\exp(-2|\Delta)\G)$ in both phases, in the same way as the other irrelevant variable, $\av{\ln(\W/J^x)}$ in (a), and  $\av{\ln(\W/h)}$ in (b).}
 \label{fig:off_critical_flow}
\end{figure}

To find the scaling of the interactions off criticality and make sure they are irrelevant we apply the RG steps numerically on a chain with $\sim10^6$ sites. The results of the flow for both phases are shown in Fig. \ref{fig:off_critical_flow}. We find that $\av{\ln(\W/J^x)}\sim\exp(-2|\Delta)\G)$ in both phases. Note that in each phase there is another variable with the same scaling, $\av{\ln(\W/J^x)}$ in the paramagnetic phase, and $\av{\ln(\W/h)}$ in the spin-glass phase. However, since the pre-factor of the exponent is larger for the flow of $J^x$, we get that both $\av{\ln(J^z/J^x)}$ and $\av{\ln(h/J^x)}$ grow exponentially to infinity as $\sim\exp(2|\Delta)\G)$. This result implies that all the approximations regarding the smallness of the interaction term are asymptotically exact.

\end{document}